\documentclass[12pt]{article}
\usepackage{amssymb, amsmath, graphicx}

\def\R{\mathbb{R}}
\def\C{\mathbb{C}}
\def\N{\mathbb{N}}

\def\I{{\rm 1\kern -.3600em 1}}

\newtheorem{theorem}{Theorem}%[section]
\newtheorem{corollary}[theorem]{Corollary}
\newtheorem{lemma}[theorem]{Lemma}
\newtheorem{proposition}[theorem]{Proposition}
\newtheorem{definition}[theorem]{Definition}

\begin{document}

\author{\textbf{Cust\'odia Drumond} \\
{\small DME, University of Madeira, P 9000-390 Funchal, Portugal}\\
{\small CCM, University of Madeira, P 9000-390 Funchal, Portugal}\\
{\small collie@uma.pt} \and
\textbf{Maria Jo\~{a}o Oliveira} \\
{\small Universidade Aberta, P 1269-001 Lisbon, Portugal}\\
{\small CMAF, University of Lisbon, P 1649-003 Lisbon, Portugal}\\
{\small oliveira@cii.fc.ul.pt} \and
\textbf{Jos\'e Lu{\'\i}s da Silva} \\
{\small DME, University of Madeira, P 9000-390 Funchal, Portugal}\\
{\small CCM, University of Madeira, P 9000-390 Funchal, Portugal}\\
{\small luis@uma.pt}}

\title{Intersection local times of fractional Brownian motions with 
$H\in\left(0,1\right)$ as generalized white noise functionals}

\date{}
\maketitle

\vspace*{-0.5cm}

\begin{abstract}
In $\R^d$, for any dimension $d\geq 1$, expansions of self-intersection local 
times of fractional Brownian motions with arbitrary Hurst coefficients in 
$(0,1)$ are presented. The expansions are in terms of Wick powers of white 
noises (corresponding to multiple Wiener integrals), being well-defined in the 
sense of generalized white noise functionals.
\end{abstract}

Dedicated to Professor Ludwig Streit on the occasion of his $70^{\textrm{th}}$ 
Birthday

\bigskip


\medskip

\noindent
\textbf{2000 AMS Classification:} 60H40, 60G15, 28C20, 46F25

\newpage

\section{Introduction}

%\usepackage{mathptmx}
%\usepackage[T1]{fontenc}
%\usepackage[latin9]{inputenc}
%\usepackage{esint}

%\makeatletter

%\providecommand{\boldsymbol}[1]{\mbox{\boldmath $#1$}}

%\usepackage{babel}
%\makeatother

Intersection local times of fractional Brownian motion (fBm) has been
studied by many authors, see e.g.~the works done by Gradinaru et 
al.~\cite{GRV03}, Nualart et al.~\cite{NH07}, \cite{NH05}, Rosen \cite{R87}, 
and the references therein. 

We may consider intersections of simple paths of a fBm with themselves
or with other independent fractional Brownian motion as in \cite{NL07}.
Within the white noise analysis framework, we concentrate here in the 
simple self-intersection problem using a similar approach to Faria et 
al.~\cite{FHSW97}, \cite{FDS00} to study the case $H=1/2$ (the classical
Brownian motion). This approach has the advantage that the underlying 
probability space does not depend on the Hurst coefficient under consideration.
As a consequence, we analyze the self-intersections of a fBm, with Hurst 
coefficient being any possible value in $(0,1)$. As one may expect, 
independently of the Hurst coefficient under consideration, self-intersections 
become scarce as the dimension $d$ increases. 

An informal but suggestive definition of self-intersection local time
of a generic $d$-dimensional fBm $B$ is in terms of an integral over Donsker's 
$\delta$-function
\[
L=\int d^{2}t\,\delta(B(t_{2})-B(t_{1})),
\]
intended to measure the amount of ``time'' the process spends
intersecting itself. 

A rigorous definition, such as, e.g., through a sequence of Gaussians 
approximating the $\delta$-function,
\[
(2\pi\varepsilon)^{-d/2}\exp\left(-\frac{|x|^{2}}{2\varepsilon}\right),\quad\varepsilon>0,
\]
will make $L$ increasingly singular, and various ``renormalizations'' 
have to be done as the dimension $d$ increases. For $d>1$ and $1/d\leq H<3/(2+d)$, 
the expectation diverges in the limit and must be subtracted. The 
$L^2$-properties related to this situation have been analyzed in \cite{NH05}. 
For other $H$ values further kernel terms must be subtracted (Theorem 
\ref{Theorem1} below).

In this work we are particularly interested in the chaos decomposition of $L$. 
We expand $L$ in terms of Wick powers \cite{HKPS93} of white noise,
an expansion which corresponds to that in terms of multiple Wiener integrals
when one considers the Wiener process as the fundamental random variable.
This allows us to derive the kernels for $L$. Due to the local structure of the 
Wick powers, the kernel functions are relatively simple and exhibit clearly 
the dimension dependence singularities of $L$ (Proposition \ref{Proposition2}).
For comparison, we also calculate the regularized kernel functions
corresponding to the Gaussian $\delta$-sequence mentioned above.

The paper is organized as follows. In Section~2 we review the necessary
background of white noise analysis and in Section~3 we present the
main results of this work and their proofs. We shall mention that after the
conclusion of our work we found the recent paper \cite{Re07} with related 
results.

\section{Gaussian white noise calculus}

In this section we briefly recall the concepts and results of white
noise analysis used throughout this work (for a detailed explanation see 
e.g.~\cite{BeKo88}, \cite{Hid75}, \cite{HKPS93}, \cite{HOUZ96}, \cite{Ko75}, 
\cite{KuTa80a}, \cite{KuTa80}, \cite{Kuo96}, \cite{Ob94}). 

\subsection{Fractional Brownian motion}

The starting point of white noise analysis is the real Gelfand triple 
\[
S_d(\R)\subset L_d^2(\R)\subset S_d'(\R),
\]
where $L_d^{2}(\R):=L^2(\R,\R^d)$, $d\geq 1$, is the real Hilbert space of all 
vector valued square integrable functions with respect to the Lebesgue 
measure on $\R$, $S_d(\R)$ and $S_d'(\R)$ are the Schwartz spaces of the 
vector valued test functions and tempered distributions, respectively. We shall
denote the $L^2_d(\R)$-norm by $|\cdot|$ and the dual pairing between 
$S_d'(\R)$ and $S_d(\R)$ by $\left\langle \cdot,\cdot\right\rangle$, 
which is defined as the bilinear extension of the inner product on 
$L_d^2(\R)$, i.e.,
\[
\langle\mathbf{g},\mathbf{f}\rangle= \sum_{i=1}^d \int_{\R} dx\,g_i(x)f_i(x), 
\]
for all $\mathbf{g}=(g_1,..., g_d)\in L^2_d(\R)$ and all
$\mathbf{f}=(f_1,..., f_d)\in S_d(\R)$. By the Minlos theorem, there is
a unique probability measure $\mu$ on the $\sigma$-algebra $\mathcal{B}$ 
generated by the cylinder sets on $S'_d(\R)$ with characteristic function 
given by
\[
C(\mathbf{f}):=\int_{S'_d(\R)}d\mu(\mathbf{\omega})\,e^{i\left\langle\mathbf{\omega},\mathbf{f}\right\rangle }
=e^{-\frac{1}{2}|\mathbf{f}|^{2}},\quad\mathbf{f}\in S_d(\R).
\]
In this way we have defined the white noise measure space 
$(S_d'(\R),\mathcal{B},\mu)$. Within this formalism, a version of the 
$d$-dimensional Wiener Brownian motion is given by 
\[
\mathbf{B}(t):=\left(\langle\omega_1,\I_{[0,t]}\rangle,...,
\langle\omega_d,\I_{[0,t]}\rangle\right),\quad (\omega_1,...,\omega_d)\in S_d'(\R)
\]
where $\I_A$ denotes the indicator function of a set $A$. For an 
arbitrary Hurst parameter $0<H<1$, $H\not=\frac12$, a version of a 
$d$-dimensional fractional Brownian motion is given by
\[
\mathbf{B}_H(t):=\left(\langle\omega_1,M_H\I_{[0,t]}\rangle,...,
\langle\omega_d,M_H\I_{[0,t]}\rangle\right),\quad (\omega_1,...,\omega_d)\in S_d'(\R),
\]
where, for a generic real valued function $f$, and for $\frac12<H<1$,
\begin{equation}
(M_Hf)(x):=\frac{K_H}{\Gamma\left(H-\frac12\right)}
\int_x^{\infty}dy\,f(y)(y-x)^{H-\frac32},\label{08eq1} 
\end{equation}
provided the integral exists for all $x\in\R$, while for $0<H<\frac12$,
\begin{equation}
(M_Hf)(x):=\frac{(\frac12-H)K_H}{\Gamma\left(H+\frac12\right)}\lim_{\varepsilon\to 0^+}\int_\varepsilon^{\infty}dy\,\frac{f(x)-f(x+y)}{y^{\frac32-H}},\label{08eq2}
\end{equation}
provided the limit exists for almost all $x\in\R$ (for more details see 
e.g.~\cite{B03} and \cite{PT00} and the references therein). Independently of 
the case under consideration, the normalizing constant $K_H$ is given by
\[
K_H=\Gamma\left(H+\frac12\right)
\left(\frac{1}{2H}+\int_0^{\infty}ds\,\left((1+s)^{H-\frac12}-s^{H-\frac12}\right)\right)^{-\frac12}.
\]

Apart from the operators (\ref{08eq1}), (\ref{08eq2}), we shall also consider 
the operator defined for $\frac12<H<1$ by
\begin{equation}
(M^+_Hf)(x):=\frac{K_H}{\Gamma\left(H-\frac12\right)}
\int_{-\infty}^xdy\,f(y)(x-y)^{H-\frac32},\label{08eq3}
\end{equation}
provided the integral exists for all $x\in\R$, and the operator defined for 
$0<H<\frac12$ by
\begin{equation}
(M^+_Hf)(x):=\frac{(\frac12-H)K_H}{\Gamma\left(H+\frac12\right)}\lim_{\varepsilon\to 0^+}\int_\varepsilon^{\infty}dy\,\frac{f(x)-f(x-y)}{y^{\frac32-H}},\label{08eq4}
\end{equation}
provided the limit exists for almost all $x\in\R$. 

There are several examples of functions $f$ for which $M_Hf$ and $M_H^+f$ 
exist for any $H\in(0,1)$. For instance, $f=\I_{[0,t]}$, $t>0$, or 
$f\in S_1(\R)$. For functions $f_1$, $f_2$ being either one of these two types 
it is easy to prove the following equality
\[
\int_{\R}dx\,f_1(x)(M_Hf_2)(x)=\int_{\R}dx\,(M_H^+f_1)(x)f_2(x),
\]
showing that $M_H$ and $M_H^+$ are dual operators. For more details and proofs 
see e.g.~\cite{B03} and the references therein.   
        
\subsection{Hida distributions and characterization results}

Let us now consider the complex Hilbert space 
$L^{2}(S'_d(\R),\mathcal{B},\mu)$. This space is canonically 
isomorphic to the symmetric Fock space of symmetric square integrable 
functions,
$$
L^{2}(S'_d(\R),\mathcal{B},\mu)\simeq 
\Big(\bigoplus_{k = 0}^\infty \mathrm{Sym}\, L^2(\R^k, k!d^kx)\Big)^{\otimes d},
$$
leading to the chaos expansion of the elements in 
$L^{2}(S'_d(\R),\mathcal{B},\mu)$, 
\begin{equation} 
F(\omega_1,...,\omega_d) = \sum_{(n_1,...,n_d)\in\N^d}
\langle:\omega_1^{\otimes n_1}: \otimes\cdots\otimes 
: \omega_d^{\otimes n_d}:,\mathbf{f}_{(n_1,...,n_d)}\rangle,\label{08eq5}
\end{equation}
with kernel functions $\mathbf{f}_{(n_1,...,n_d)}$ in the Fock space. For 
simplicity, in the sequel we shall use the notation
\[
\mathbf{n} =(n_1, \cdots ,n_d)\in \N^d,\quad n = \sum_{i = 1}^d n_i,\quad 
\mathbf{n}! = \prod_{i = 1}^d n_i!,
\]
which reduces expansion (\ref{08eq5}) to
\[
F(\mathbf{\omega})= \sum_{\mathbf{n}\in\N^d}\langle:\mathbf{\omega}^{\otimes\mathbf{n}}:, 
\mathbf{f}_{\mathbf{n}}\rangle,\quad \mathbf{\omega}\in S'_d(\R).
\]

To proceed further we have to consider a Gelfand triple around the space 
$L^{2}(S'_d(\R),\mathcal{B},\mu)$. We will use the space $(S)^*$ of Hida 
distributions (or generalized Brownian functionals) and the corresponding 
Gelfand triple $(S)\subset L^{2}(S'_d(\R),\mathcal{B},\mu)\subset(S)^*$. Here 
$(S)$ is the space of white noise test functions such that its dual space 
(with respect to $L^{2}(S'_d(\R),\mathcal{B},\mu)$) is the space $(S)^*$. 
Instead of 
reproducing the explicit construction of $\left(S\right)^*$ (see
e.g.~\cite{HKPS93}), in Theorem \ref{08Prop1} below we define this space by its 
$S$-transform. We recall that given a $\mathbf{f}\in S_d(\R)$, and the 
Wick exponential
\[
:\exp(\langle\mathbf{\omega},\mathbf{f}\rangle):\,:=
\sum_{\mathbf{n}\in\N^d}\frac{1}{\mathbf{n}!}
\langle:\mathbf{\omega}^{\otimes\mathbf{n}}:,\mathbf{f}^{\otimes\mathbf{n}}\rangle
=C(\mathbf{f})e^{\langle\mathbf{\omega},\mathbf{f}\rangle},
\]
we define the $S$-transform of a $\Phi\in \left(S\right)^*$ by
\begin{equation}
S\Phi(\mathbf{f}):= 
\left\langle\!\left\langle \Phi,:\exp(\left\langle \cdot,\mathbf{f}
\right\rangle):\right\rangle\!\right\rangle,\quad \forall\,\mathbf{f}\in S_d(\R).\label{08eq6}
\end{equation}
Here $\left\langle\!\left\langle\cdot ,\cdot\right\rangle\!\right\rangle$ 
denotes the dual pairing between $\left(S\right)^*$ and 
$\left(S\right)$ which is defined as the bilinear extension of 
the sesquilinear inner product on $L^2(S'_d(\R),\mathcal{B},\mu)$. We observe 
that the multilinear expansion of (\ref{08eq6}),
\[
S\Phi(\mathbf{f}):= \sum_{\mathbf{n}}\langle F_{\mathbf{n}},\mathbf{f}^{\otimes\mathbf{n}}\rangle,
\] 
extends the chaos expansion to $\Phi\in \left(S\right)^*$ with distribution valued kernels $F_{\mathbf{n}}$ such that
\[
\left\langle\!\left\langle\Phi,\varphi\right\rangle\!\right\rangle
=\sum_{\mathbf{n}}\mathbf{n}!\langle F_{\mathbf{n}},\varphi_{\mathbf{n}}\rangle,
\]
for every generalized test function $\varphi\in(S)$ with kernel functions 
$\varphi_{\mathbf{n}}$.

In order to characterize the space $\left(S\right)^*$ through its 
$S$-transform we need the following definition.

\begin{definition}
\label{Def1}A function $F:S_d(\R)\rightarrow \C$ is called a 
$U$-functional whenever\newline
1. for every $\mathbf{f}_1,\mathbf{f}_2\in S_d(\R)$ the mapping 
$\Bbb{R\ni \lambda }\longmapsto F(\lambda \mathbf{f}_1+\mathbf{f}_2)$ has an 
entire extension to $\lambda \in \C$,\newline
2. there are constants $K_1,K_2>0$ such that 
\[
\left| F(z\mathbf{f})\right| \leq K_1\exp \left( K_2\left| z\right| ^2\left\|
\mathbf{f}\right\| ^2\right) ,\quad \forall \,z\in \C,\mathbf{f}\in S_d(\R)
\]
for some continuous norm $\left\| \cdot \right\|$ on $S_d(\R)$.
\end{definition}

We are now ready to state the aforementioned characterization result.

\begin{theorem}
\label{08Prop1}{\rm (\cite{KLPSW96}, \cite{PS91})} The $S$-transform defines a 
bijection between the space $\left(S\right)^*$ and the space of 
$U$-functionals.
\end{theorem}

As a consequence of Theorem \ref{08Prop1} one may derive the next two 
statements. The first one concerns the convergence of sequences of 
Hida distributions and the second one the Bochner integration of families of 
the same type of distributions (for more details and proofs see 
e.g.~\cite{HKPS93}, \cite{KLPSW96}, \cite{PS91}).

\begin{corollary}
\label{Corollary2}Let $\left( \Phi _n\right)_{n\in\N}$ be a sequence in 
$\left(S\right)^*$ such that
\begin{description}     
\item[{\it (i)}] For all $\mathbf{f}\in S_d(\R)$, 
$\left((S\Phi_n)(\mathbf{f})\right)_{n\in\N}$ is a Cauchy sequence in $\C$, 
\item[{\it (ii)}] There are $K_1,K_2>0$ such that for some continuous norm 
$\left\| \cdot \right\|$ on $S_d(\R)$ one has  
\[
\left| (S\Phi_n)(z\mathbf{f})\right| \leq K_1e^{K_2\left| z\right| ^2\left\|
\mathbf{f}\right\| ^2} ,\quad \forall \,z\in \C,\mathbf{f}\in S_d(\R),n\in\N.
\]
\end{description}
Then $\left(\Phi _n\right)_{n\in\N}$ converges strongly in 
$\left(S\right)^*$ to an unique Hida distribution. 
\end{corollary}

\begin{corollary}
\label{Corollary1} Let $(\Omega, \mathcal{B}, m)$ be a measure space and 
$\lambda\mapsto \Phi_\lambda$ be a mapping from $\Omega$ to $(S)^*$. We assume that the $S$-transform of $\Phi_\lambda$ fulfills the following two properties:
\begin{description}     
\item[{\it (i)}] The mapping 
$\lambda\mapsto (S\Phi_\lambda)(\mathbf{f})$ is measurable for every $\mathbf{f}\in S_d(\R)$,
\item[{\it (ii)}] The $S\Phi_\lambda$ obeys a $U$-estimate
\[
|(S\Phi_\lambda)(z\mathbf{f})| \leq C_1(\lambda) e^{C_2(\lambda) |z|^2 
\Vert\mathbf{f}\Vert^2},\quad z\in\C,\mathbf{f}\in S_d(\R),
\]
for some continuous norm $\Vert\cdot\Vert$ on $S_d(\R)$ and for 
$C_1\in L^1(\Omega,m)$, $C_2\in L^\infty(\Omega,m)$.
\end{description}
Then 
\[
\int_\Omega dm(\lambda)\, \Phi_\lambda\in (S)^*
\]
and
\[
S\left(\int_\Omega dm(\lambda)\,\Phi_\lambda\right) (\mathbf{f}) =
\int_\Omega dm(\lambda)\,S\Phi_\lambda(\mathbf{f}).
\] 
\end{corollary}

\section{Intersection local times}

\begin{proposition} For $t\not= s$ the Bochner integral 
\begin{equation}
\delta (\mathbf{B}_H(t)-\mathbf{B}_H(s)):=\left(\frac 1{2\pi }\right)^d
\int_{\R^d}
d\mathbf{\lambda}\,e^{i\mathbf{\lambda}(\mathbf{B}_H(t)-\mathbf{B}_H(s))}\label{eq1}
\end{equation}
is a Hida distribution with $S$-transform given by
\begin{equation}
S\delta (\mathbf{B}_H(t)-\mathbf{B}_H(s))(\mathbf{f})=
\left(\frac 1{\sqrt{2\pi}|t-s|^H}\right)^d
e^{-\frac 1{2|t-s|^{2H}}\left|\int_{\R}dx\,
\mathbf{f}(x)(M_H\I_{\left[s\wedge t,s\vee t\right]})(x)\right|^2},\label{eq4}
\end{equation}
for all $\mathbf{f}\in S_d(\R)$.
\end{proposition}

\noindent
\textbf{Proof.} The proof of this result follows from the application of 
Corollary \ref{Corollary1} to the $S$-transform of the integrand function in 
(\ref{eq1}) with respect to the Lebesgue measure on $\R^d$.

Since
\begin{equation}
Se^{i\mathbf{\lambda}(\mathbf{B}_H(t)-\mathbf{B}_H(s))}(\mathbf{f})
=e^{-\frac{|\mathbf{\lambda}|^2}{2}(t-s)^{2H}}e^{i\mathbf{\lambda}\int_{\R}dx\,
\mathbf{f}(x)(M_H\I_{\left[s,t\right]})(x)},\label{eq2}
\end{equation}
for $t>s$, and 
\begin{equation}
Se^{i\mathbf{\lambda}(\mathbf{B}_H(t)-\mathbf{B}_H(s))}(\mathbf{f})
=e^{-\frac{|\mathbf{\lambda}|^2}{2}(s-t)^{2H}}e^{-i\mathbf{\lambda}
\int_{\R}dx\,\mathbf{f}(x)(M_H\I_{\left[t,s\right]})(x)},\label{eq3}
\end{equation}
for $t<s$ cf.~e.g.~\cite{HKPS93}, clearly in both situations the measurability 
condition is fulfilled. 

Independently of the particular case under consideration, we observe that for 
all $z\in\C$ and all $\mathbf{f}\in S_d(\R)$ we find
\begin{eqnarray*}
\left|Se^{i\mathbf{\lambda}(\mathbf{B}_H(t)-\mathbf{B}_H(s))}(z\mathbf{f})\right|&=&e^{-\frac{|\mathbf{\lambda}|^2}{4}|t-s|^{2H}}
\left|e^{-\frac{|\mathbf{\lambda}|^2}{4}|t-s|^{2H}
\pm iz\mathbf{\lambda}\int_{\R}dx\,
\mathbf{f}(x)(M_H\I_{\left[s\wedge t,s\vee t\right]})(x)}\right|\\
&\leq& e^{-\frac{|\mathbf{\lambda}|^2}{4}|t-s|^{2H}}
e^{-\frac{|\mathbf{\lambda}|^2}{4}|t-s|^{2H}+
|z||\mathbf{\lambda}|\left|\int_{\R}dx\,
\mathbf{f}(x)(M_H\I_{\left[s\wedge t,s\vee t\right]})(x)\right|},
\end{eqnarray*}
where the latter exponential is bounded by
\[
e^{\frac{|z|^2}{|t-s|^{2H}}\left|\int_{\R}dx\,\mathbf{f}(x)
(M_H\I_{\left[s\wedge t,s\vee t\right]})(x)\right|^2},
\]
because
\begin{eqnarray*}
&&-\frac{|\mathbf{\lambda}|^2}{4}|t-s|^{2H}+
|z||\mathbf{\lambda}|\left|\int_{\R}dx\,
\mathbf{f}(x)(M_H\I_{\left[s\wedge t,s\vee t\right]})(x)\right|\\ 
&=&-\left(\frac{|z|}{|t-s|^H}\left|
\int_{\R}dx\,\mathbf{f}(x)(M_H\I_{\left[s\wedge t,s\vee t\right]})(x)
\right|-\frac{|\mathbf{\lambda}|}{2}|t-s|^H\right)^2\\
&&+\frac{|z|^2}{|t-s|^{2H}}\left|
\int_{\R}dx\,\mathbf{f}(x)(M_H\I_{\left[s\wedge t,s\vee t\right]})(x)\right|^2.
\end{eqnarray*}
As a result,
\[
\left|Se^{i\mathbf{\lambda}(\mathbf{B}_H(t)-\mathbf{B}_H(s))}(z\mathbf{f})\right|\leq e^{-\frac{|\mathbf{\lambda}|^2}{4}|t-s|^{2H}}
e^{\frac{|z|^2}{|t-s|^{2H}}\left|\int_{\R}dx\,\mathbf{f}(x)
(M_H\I_{\left[s\wedge t,s\vee t\right]})(x)\right|^2},
\]
where, as a function of $\mathbf{\lambda}$, the first factor is integrable on 
$\R^d$ and the second factor is constant. 

An application of the result mentioned above completes the proof. In 
particular, it yields (\ref{eq4}) by integrating (\ref{eq2}), (\ref{eq3}) over 
$\mathbf{\lambda}$.\hfill$\blacksquare \medskip$

In order to proceed further the next result shows to be very useful. It
improves the estimate obtained in \cite[Theorem 2.3]{B03} towards the 
characterization results stated in Corollaries \ref{Corollary2} and 
\ref{Corollary1}.

\begin{lemma}
\label{Lemma} Let $H\in (0,1)$ and $f\in S_1(\R)$ be given. 
There is a non-negative constant $C_H$ independent of $f$ such that
\[
\left|\int_{\R}dx\,f(x)(M_H\I_{\left[s,t\right]})(x)\right|
\leq C_H|t-s|\left(\sup_{x\in\R}|f(x)|+ \sup_{x\in\R}|f'(x)|
+|f|\right)
\]
for all $s<t$.
\end{lemma}

\noindent
\textbf{Proof.} Since $M_H$ and $M_H^+$ are dual operators, 
\[
\int_{\R}dx\,f(x)(M_H\I_{\left[s,t\right]})(x)=
\int_s^t dx\,(M_H^+f)(x),
\]
and thus 
\[
\left|\int_{\R}dx\,f(x)(M_H\I_{\left[s,t\right]})(x)\right|
\leq |t-s|\sup_{x\in\R}|(M_H^+f)(x)|.
\]
It has been shown in \cite[Proof of Theorem 2.3]{B03} that for $0<H<\frac12$,
\[
\sup_{x\in\R}|(M_H^+f)(x)|\leq 
C'_H\left(\frac2{\frac12 - H}\sup_{x\in\R}|f(x)| 
+ \frac1{H+\frac12}\sup_{x\in\R}|f'(x)|\right),
\]
for some constant $C'_H>0$ independent of $f$, while for $\frac12<H<1$,
\begin{eqnarray*}
&&|(M_H^+f)(x)|\\
&\leq&C_H''\int_{\R}dy\,|f(y)||y-x|^{H-\frac32}\\
&=& C_H''\int_{|y-x|<1}dy\,|f(y)||y-x|^{H-\frac32}
+C_H''\int_{|y-x|>1}dy\,|f(y)||y-x|^{H-\frac32}\\
&\leq&C_H''\frac2{H-\frac12}\sup_{x\in\R}|f(x)|+
C_H''\int_{|y-x|>1}dy\,|f(y)||y-x|^{H-\frac32},
\end{eqnarray*}
for some constant $C_H''>0$ also independent of $f$. Concerning the latter 
integral, observe that 
\begin{eqnarray*}
\int_{|y-x|>1}dy\,|f(y)||y-x|^{H-\frac32}
&\leq&|f|\left(\int_{|y-x|>1}dy\,|y-x|^{2H-3}\right)^{1/2}\\
&=&\frac1{\sqrt{1-H}}|f|,
\end{eqnarray*}
leading to
\[
\sup_{x\in\R}|(M_H^+f)(x)|\leq C_H''\left(\frac2{H-\frac12}\sup_{x\in\R}|f(x)|+
\frac1{\sqrt{1-H}}|f|\right).
\]
\hfill$\blacksquare \medskip$

We are now ready to state the main result on intersection local times $L_H$ as 
well as on their subtracted counterparts $L_H^{(N)}$. For simplicity we shall 
use the notation
\[
\Delta:=\{(t_1,t_2)\in\R^2: 0<t_1<t_2<1\}.
\]

\begin{theorem}
\label{Theorem1} For any $H\in (0,1)$ and for any pair of integer numbers 
$d\geq 1$ and $N\geq 0$ such that $2N(H-1)+dH<1$, the Bochner integral
\[
L_H^{(N)}:=\int_\Delta d^2t\,\delta^{(N)}(\mathbf{B}_H(t_2)-\mathbf{B}_H(t_1))
\]
is a Hida distribution.
\end{theorem}

\noindent
\textbf{Proof.}  To prove this result we shall again use Corollary 
\ref{Corollary1} with respect to the Lebesgue measure on $\Delta$. For this 
purpose let us denote the truncated exponential series by
\[
\exp_N (x) := \sum_{n = N}^\infty {{x^n}\over {n!}}.
\] 
It follows from (\ref{eq4}) that the $S$-transform of $\delta^{(N)}$ is given 
by 
\begin{eqnarray}
&&S(\delta^{(N)}(\mathbf{B}_H(t_2)-\mathbf{B}_H(t_1)))(\mathbf{f})\label{eq6}\\
&=&\!\!\!\!\!\left(\frac 1{\sqrt{2\pi}|t_2-t_1|^H}\right)^d
\!\!\exp_N\left(-\frac 1{2|t_2-t_1|^{2H}}\left|\int_{\R}dx\,
\mathbf{f}(x)(M_H\I_{\left[t_1,t_2\right]})(x)\right|^2\right),\nonumber
\end{eqnarray}
which is a measurable function.

In order to check the boundedness condition, on $S_d(\R)$ let us consider the 
norm $\Vert\cdot\Vert$ defined by
\begin{equation}
\Vert\mathbf{f}\Vert:= \left(\sum_{i=1}^d\left(\sup_{x\in\R}|f_i(x)|
+ \sup_{x\in\R}|f_i'(x)| + |f_i|\right)^2\right)^{\frac12},\quad \mathbf{f}=(f_1,...,f_d)\in S_d(\R).\label{08equ1}
\end{equation}
We observe that for $d=1$ this norm reduces to a continuous norm on $S_1(\R)$,
\[
\Vert f\Vert=\sup_{x\in\R}|f(x)|+\sup_{x\in\R}|f'(x)| + |f|,\quad f\in S_1(\R),
\]
which implies the continuity of the norm (\ref{08equ1}) for higher dimensions. 
By Lemma \ref{Lemma} we obtain
\[
\left|\int_{\R}dx\,\mathbf{f}(x)(M_H\I_{\left[t_1,t_2\right]})(x)\right|^2\leq 
C_H^2|t_2-t_1|^2\Vert\mathbf{f}\Vert^2,
\]
and thus
\begin{eqnarray*}
&&\left|S(\delta^{(N)}(\mathbf{B}_H(t_2)-\mathbf{B}_H(t_1)))(z\mathbf{f})
\right|\\
&\leq&\left(\frac 1{\sqrt{2\pi}|t_2-t_1|^H}\right)^d
\exp_N\left(\frac{C_H^2}2|z|^2|t_2-t_1|^{2-2H}\Vert\mathbf{f}\Vert^2\right),
\end{eqnarray*}
for all $z\in\C$ and all $\mathbf{f}\in S_d(\R)$. Estimating the function 
$\exp_N$ by
\[
\exp_N\left(\frac{C_H^2}2|z|^2|t_2-t_1|^{2-2H}\Vert\mathbf{f}\Vert^2\right)
\leq |t_2-t_1|^{2N(1-H)}e^{\frac{C_H^2}{2}|z|^2\Vert\mathbf{f}\Vert^2},
\] 
we then obtain
\[
\left|S(\delta^{(N)}(\mathbf{B}_H(t_2)-\mathbf{B}_H(t_1)))(z\mathbf{f})
\right|
\leq\left(\frac1{\sqrt{2\pi}}\right)^d |t_2-t_1|^{2N(1-H)-dH}
e^{\frac{C_H^2}2|z|^2\Vert\mathbf{f}\Vert^2},
\]
where $|t_2-t_1|^{2N(1-H)-dH}$ is integrable on $\Delta$ if, and only if, 
$2N(1-H)-dH>-1$. The proof is completed by an application of Corollary 
\ref{Corollary1}.\hfill$\blacksquare \medskip$

As a consequence, one may derive the chaos expansion for the (truncated) local 
times $L_H^{(N)}$.

\begin{proposition}
\label{Proposition2} 
Given a $H\in (0,1)$ and a pair of integer numbers $d\geq 1$ and $N\geq 0$ 
such that $2N(H-1)+dH<1$, the kernel functions $F_{H,\mathbf{n}}$ are given by
\[
F_{H,2\mathbf{n}}(u_1,...,u_{2n})=\frac1{\mathbf{n}!}
\left(\frac 1{\sqrt{2\pi}}\right)^d\left(-\frac12\right)^n\int_\Delta d^2t\,\frac 1{|t_2-t_1|^{2Hn+dH}}\prod_{i=1}^{2n}(M_H\I_{\left[t_1,t_2\right]})(u_i)
\] 
for each $\mathbf{n}\in\N^d$ such that $n\geq N$. All other kernel functions
$F_{H,\mathbf{n}}$ are identically equal to zero.
\end{proposition}

\noindent
\textbf{Proof.} By Corollary \ref{Corollary1}, the $S$-transform of the 
(truncated) local time $L_H^{(N)}$ is obtained by integrating (\ref{eq6}) over 
$\Delta$. Hence, given a $\mathbf{f}=(f_1,...,f_d)\in S_d(\R)$, 
\begin{eqnarray*}
SL_H^{(N)}(\mathbf{f})&=& \left(\frac 1{\sqrt{2\pi}}\right)^d
\int_\Delta d^2t\,\frac 1{|t_2-t_1|^{dH}}\sum_{n=N}^\infty
\frac{(-1)^n}{2^n|t_2-t_1|^{2Hn}}\times\\
&&\times\sum_{{n_1,\cdots ,n_d}\atop{n_1 +\cdots + n_d = n}}
\frac 1{n_1!\cdots n_d!}
\prod_{j=1}^d\left(\int_{\R}dx\,f_j(x)(M_H\I_{\left[t_1,t_2\right]})(x)\right)^{2n_j}\\
&=&\left(\frac 1{\sqrt{2\pi}}\right)^d
\int_\Delta d^2t \sum_{n=N}^\infty
\left(-\frac12\right)^n\frac 1{|t_2-t_1|^{2Hn+dH}}\times\\
&&\times\sum_{{n_1,\cdots ,n_d}\atop{n_1 +\cdots + n_d = n}}
\frac1{\mathbf{n}!}
\prod_{j=1}^d\left(\int_{\R}dx\,f_j(x)(M_H\I_{\left[t_1,t_2\right]})(x)\right)^{2n_j}.
\end{eqnarray*}
Comparing with the general form of the chaos expansion
\[
L_H^{(N)} = \sum_{\mathbf{n}}\langle:\mathbf{\omega}^{\otimes\mathbf{n}}:, 
F_{H,\mathbf{n}}\rangle,
\] 
one concludes that 
\[
F_{H,2\mathbf{n}}(u_1,...,u_{2n})=\frac1{\mathbf{n}!}
\left(\frac 1{\sqrt{2\pi}}\right)^d\left(-\frac12\right)^n
\int_\Delta d^2t\,\frac 1{|t_2-t_1|^{2Hn+dH}}\prod_{i=1}^{2n}(M_H\I_{\left[t_1,t_2\right]})(u_i)
\]
for each $\mathbf{n}\in\N^d$ such that $n\geq N$, while all other kernels 
$F_{H,\mathbf{n}}$ vanish.\hfill$\blacksquare \medskip$

Theorem \ref{Theorem1} shows that for $d=1$ all intersection local times $L_H$ 
are well-defined for all possible Hurst parameters $H$ in $(0,1)$. For 
$d\geq 2$, intersection local times are well-defined only for $H<1/d$. 
Informally speaking, 
for $H\geq 1/d$ with $d\geq 2$, the local times only become well-defined once 
subtracted the divergent terms. This ``renormalization'' procedure motivates 
the study of a regularization. As a computationally simple regularization we 
discuss
\[
L_{H,\varepsilon}:=
\int_\Delta d^2t\,
\delta_\varepsilon(\mathbf{B}_H(t_2)-\mathbf{B}_H(t_1)),\quad\varepsilon >0,
\] 
where
\[
\delta_\varepsilon(\mathbf{B}_H(t_2)-\mathbf{B}_H(t_1)):=
\left(\frac1{\sqrt{2\pi\varepsilon}}\right)^d 
e^{-\frac{(\mathbf{B}_H(t_2)-\mathbf{B}_H(t_1))^2}{2\varepsilon}}.
\]
\begin{theorem}
Let $\varepsilon >0$ be given. For all $H\in (0,1)$ and all dimensions $d\geq 1$
the functional $L_{H,\varepsilon}$ is a Hida distribution with kernel 
functions given by
\begin{eqnarray*}
&&F_{H,\varepsilon, 2\mathbf{n}}(u_1,...,u_{2n})\\
&=&
\frac1{\mathbf{n}!}\left(\frac 1{\sqrt{2\pi}}\right)^d\left(-\frac12\right)^n
\int_\Delta d^2t\,\frac 1{(\varepsilon+|t_2-t_1|^{2H})^{n+\frac{d}{2}}}\prod_{i=1}^{2n}(M_H\I_{\left[t_1,t_2\right]})(u_i)
\end{eqnarray*}
for each $\mathbf{n}=(n_1,...,n_d)\in\N^d$, and 
$F_{H,\varepsilon, \mathbf{n}}\equiv 0$ if at least one of the $n_i$ is an odd 
number. Moreover, if $2N(H-1)+dH<1$, then when $\varepsilon$ goes to zero the 
(truncated) functional $L^{(N)}_{H,\varepsilon}$ converges strongly in $(S)^*$ 
to the (truncated) local time $L_H^{(N)}$.  
\end{theorem}

\noindent
\textbf{Proof.} As before, the first part of the proof follows from the 
Corollary \ref{Corollary1} with respect to the Lebesgue measure on $\Delta$. 
By the definition of the $S$-transform, for all $\mathbf{f}\in S_d(\R)$ one 
finds
\begin{eqnarray*}
&&S\delta_\varepsilon(\mathbf{B}_H(t_2)-\mathbf{B}_H(t_1))(\mathbf{f})\\
&=&\left(\frac1{\sqrt{2\pi(\varepsilon+|t_2-t_1|^{2H})}}\right)^d
e^{-\frac1{2(\varepsilon+|t_2-t_1|^{2H})}\left|\int_{\R}dx\,
\mathbf{f}(x)(M_H\I_{\left[t_1,t_2\right]})(x)\right|^2},
\end{eqnarray*}
which is measurable. Hence, by Lemma \ref{Lemma}, for all $z\in\C$ and all 
$\mathbf{f}\in S_d(\R)$
one has  
\[
\left|S\delta_\varepsilon(\mathbf{B}_H(t_2)-\mathbf{B}_H(t_1))(z\mathbf{f})\right|
\leq\left(\frac1{\sqrt{2\pi(\varepsilon+|t_2-t_1|^{2H})}}\right)^d
e^{C_H^2|z|^2\frac{|t_2-t_1|^2}{2(\varepsilon+|t_2-t_1|^{2H})}\Vert\mathbf{f}\Vert^2},
\]
with $\frac{|t_2-t_1|^2}{2(\varepsilon+|t_2-t_1|^{2H})}$ bounded on $\Delta$ 
and $(\varepsilon+|t_2-t_1|^{2H})^{-\frac{d}{2}}$ integrable on $\Delta$. By 
Corollary \ref{Corollary1}, one may then conclude that 
$L_{H,\varepsilon}\in (S)^*$ and, moreover, for every 
$\mathbf{f}=(f_1,...,f_n)\in S_d(\R)$,
\begin{eqnarray*}
SL_{H,\varepsilon}(\mathbf{f})\!\!&=&\!\!\int_\Delta d^2t\,S\delta_\varepsilon(\mathbf{B}_H(t_2)-\mathbf{B}_H(t_1))(\mathbf{f})\\
&=&\!\!\left(\frac 1{\sqrt{2\pi}}\right)^d
\int_\Delta d^2t\,\frac 1{(\varepsilon+|t_2-t_1|^{2H})^{\frac{d}{2}}}\sum_{n=0}^\infty
\frac{(-1)^n}{2^n(\varepsilon+|t_2-t_1|^{2H})^n}\times\\
&&\times\sum_{{n_1,\cdots ,n_d}\atop{n_1 +\cdots + n_d = n}}
\frac 1{n_1!\cdots n_d!}
\prod_{j=1}^d\left(\int_{\R}dx\,f_j(x)(M_H\I_{\left[t_1,t_2\right]})(x)\right)^{2n_j}\\
&=&\!\!\left(\frac 1{\sqrt{2\pi}}\right)^d
\int_\Delta d^2t \sum_{n=0}^\infty
\left(-\frac12\right)^n\frac 1{(\varepsilon+|t_2-t_1|^{2H})^{n+\frac{d}{2}}}\times\\
&&\times\sum_{{n_1,\cdots ,n_d}\atop{n_1 +\cdots + n_d = n}}
\frac1{\mathbf{n}!}
\prod_{j=1}^d\left(\int_{\R}dx\,f_j(x)(M_H\I_{\left[t_1,t_2\right]})(x)\right)^{2n_j}.
\end{eqnarray*}
As in the proof of Proposition \ref{Proposition2}, it follows from the latter 
expression that the kernels $F_{H,\varepsilon, \mathbf{n}}$ appearing in the 
chaos expansion $L_{H,\varepsilon}= \sum_{\mathbf{n}}\langle:\mathbf{\omega}^{\otimes\mathbf{n}}:, F_{H,\varepsilon,\mathbf{n}}\rangle$ vanish if at least one 
of the $n_i$ in $\mathbf{n}=(n_1,...,n_d)$ is an odd number, otherwise they 
are given by
\begin{eqnarray*}
&&F_{H,\varepsilon, 2\mathbf{n}}(u_1,...,u_{2n})\\
&=&
\frac1{\mathbf{n}!}\left(\frac 1{\sqrt{2\pi}}\right)^d\left(-\frac12\right)^n
\int_\Delta d^2t\,\frac 1{(\varepsilon+|t_2-t_1|^{2H})^{n+\frac{d}{2}}}\prod_{i=1}^{2n}(M_H\I_{\left[t_1,t_2\right]})(u_i).
\end{eqnarray*}
To complete the proof amounts to check the convergence. For this purpose we 
shall use Corollary \ref{Corollary2}. Since
\[
SL^{(N)}_{H,\varepsilon}(\mathbf{f})=\int_\Delta d^2t\,S\delta^{(N)}_\varepsilon(\mathbf{B}_H(t_2)-\mathbf{B}_H(t_1))(\mathbf{f}),
\]
for all $z\in\C$ and all $\mathbf{f}\in S_d(\R)$ we have
\begin{eqnarray*}
\left|SL^{(N)}_{H,\varepsilon}(z\mathbf{f})\right|&\leq&
\int_\Delta d^2t\,\left|S\delta_\varepsilon(\mathbf{B}_H(t_2)-\mathbf{B}_H(t_1))(z\mathbf{f})\right|\\
&\leq&\left(\frac1{\sqrt{2\pi\varepsilon}}\right)^d\int_\Delta d^2t\,
e^{\frac{C_H^2}{2\varepsilon}|z|^2|t_2-t_1|^2\Vert\mathbf{f}\Vert^2}
\leq\left(\frac1{\sqrt{2\pi\varepsilon}}\right)^d
e^{\frac{C_H^2}{2\varepsilon}|z|^2\Vert\mathbf{f}\Vert^2},
\end{eqnarray*}
showing the boundedness condition. In addition, similar computations used to 
prove Theorem \ref{Theorem1} yield for all $(t_1,t_2)\in\Delta$
\begin{eqnarray*}
&&\left|S\delta^{(N)}_\varepsilon(\mathbf{B}_H(t_2)-\mathbf{B}_H(t_1))(\mathbf{f})\right|\\
&\leq&\left(\frac1{\sqrt{2\pi}|t_2-t_1|^H}\right)^d
\exp_N\left({\frac{C_H^2}{2}|t_2-t_1|^{2-2H}\Vert\mathbf{f}\Vert^2}\right)\\
&\leq&\left(\frac1{\sqrt{2\pi}}\right)^d |t_2-t_1|^{2N(1-H)-dH}
e^{\frac{C_H^2}2\Vert\mathbf{f}\Vert^2}
\end{eqnarray*}
which allows the use of the Lebesgue dominated convergence theorem to infer 
the other condition needed for the application of Corollary \ref{Corollary2}.\hfill$\blacksquare \medskip$

\subsection*{Acknowledgments}

M.J.O.~and J.L.S.~would like to express their gratitude for the splendid
hospitality of our colleagues and friends Victoria Bernido and Christopher 
Bernido during a very pleasant stay at Jagna during the $5^{\mathrm{th}}$ Jagna 
International Workshop. Financial support of the FCT through POCI, PDCT, and 
PTDC projects is gratefully acknowledged.

%\addcontentsline{toc}{section}{References}
%\bibliographystyle{alpha}
%\bibliography{/home/oliveira/artigos/Oliveira}

\newcommand{\etalchar}[1]{$^{#1}$}

\end{document}